\documentclass[aps,twocolumn,prd,nofootinbib,superscriptaddress,preprintnumbers,10pt]{revtex4-2}

\usepackage[utf8]{inputenc}

\usepackage{mathtools}
\usepackage{amsfonts}
\usepackage{amssymb}
\usepackage{mathrsfs}
\usepackage{bbm}
\usepackage{braket}
\usepackage{slashed}
\usepackage{tensor}

\usepackage{graphicx}
\usepackage{array}

\usepackage{placeins}
\usepackage{makecell}
\usepackage[caption=false]{subfig}
\usepackage{float}

\usepackage{xspace}
\usepackage{xfrac}
\usepackage{hyperref}
\usepackage[nameinlink]{cleveref}
\usepackage{appendix}
\usepackage{units}

\usepackage{xifthen}
\usepackage{booktabs}
\usepackage{multirow}
\usepackage{courier} 
\usepackage[dvipsnames]{xcolor}
\hypersetup{
	colorlinks,
	linkcolor={blue!75!black},
	citecolor={blue!75!black},
	urlcolor={blue!75!black}
}

\Crefname{equation}{Eq.\!}{Eqs.\!}

\DeclareMathOperator{\Tr}{Tr}

\newcommand{\msbar}{\mathrm{\overline{MS}}}

\graphicspath{{./figs/}}

\newcommand{\gettitle}{The structure of the pion at large $N_c$}

\newcommand{\getMITAffiliation}{\affiliation{Center for Theoretical Physics -- A Leinweber Institute, Massachusetts Institute of Technology, Cambridge, MA 02139, USA}}
\newcommand{\getIAIFIAffiliation}{\affiliation{The NSF AI Institute for Artificial Intelligence and Fundamental Interactions}}

\newcommand{\getBernAffiliation}{\affiliation{Albert Einstein Center, Institute for Theoretical Physics, University of Bern, 3012 Bern, Switzerland}}

\hypersetup{
    pdftitle={\gettitle},
    pdfauthor={Abbott,Boyda,Fu,Hackett,Kanwar,Romero-López,Shanahan,Urban},
    pdfkeywords={lattice field theory}{lattice gauge theory}{normalizing flows}{machine learning},
    bookmarksopen=true,
    bookmarksopenlevel=2,
    bookmarksnumbered=true
}

\begin{document}

\preprint{MIT-CTP/6146}

\title{\gettitle}

\author{Isaac Zhu}
\getMITAffiliation
\author{Fernando~Romero-L\'opez}
\getBernAffiliation
\author{Phiala~E.~Shanahan}
\getMITAffiliation
\getIAIFIAffiliation

\preprint{}

\begin{abstract}
We compute the pion quark momentum fraction in the large-$N_c$ limit using the Twisted Eguchi-Kawai (TEK) reduced model. Gauge configurations for different values of the lattice spacing are generated with the pure-gauge TEK action, while the pion correlators are computed for different quark masses with maximally twisted mass fermions. The bare matrix element corresponding to the quark momentum fraction is obtained using the Feynman-Hellmann method, and we use perturbative renormalization factors to connect to the $\overline{\mathrm{MS}}$ scheme at $2\,\mathrm{GeV}$. A global analysis at several lattice spacings, pion masses, and values of $N_c$ yields $\langle x\rangle_q^{\overline{\mathrm{MS}}}(2\,\mathrm{GeV}) = 0.267(105)$ at the physical pion mass in the large-$N_c$ limit. This value is significantly smaller than existing $N_c=3$ determinations from phenomenology and lattice QCD, indicating that, in the large-$N_c$ limit, the pion's light-cone momentum is carried predominantly by gluons.
\end{abstract}

\maketitle

\section{Introduction}\label{sec:intro}

The confinement phenomenon in quantum chromodynamics (QCD) implies that colorless composite states, hadrons, are the only observable states in the theory. In the hadron spectrum, the pion constitutes the lightest state, with a low mass explained by chiral symmetry. The internal structure of the pion represents an important target for both theory and experiment~\cite{E615:1989bda,NA10:1985ibr,Owens:1984zj,Aurenche:1989sx,Sutton:1991ay,Gluck:1991ey,Wijesooriya:2005ir,Novikov:2020snp,H1:2010hym,ZEUS:2002gig,Barry:2018ort,Cao:2021aci,Pefkou:2021fni,Barry:2021osv,Aicher:2010cb,Westmark:2017uig,Kotz:2023pbu,Kotz:2025lio}, including first-principles approaches such as lattice QCD~\cite{Joo:2019bzr,Gao:2022iex,Alexandrou:2021mmi,Alexandrou:2020gxs,Hackett:2023nkr,Francis:2025pgf,Francis:2025rya}. 

The large-$N_c$ limit provides an interesting theoretical handle on QCD. In the 't Hooft limit~\cite{tHooft:1973alw}, with $g^2N_c$ and the number of quark flavors fixed as $N_c \to \infty$, gauge dynamics simplify while many qualitative features remain intact. For instance, meson observables (e.g. meson structure quantities) can be expanded in powers of $1/N_c$, and the leading-order large-$N_c$ theory can therefore be used as an approximation for ordinary $N_c=3$ QCD~\cite{Witten:1979kh,Coleman:1980nk}. In addition, the large $N_c$ limit provides a theoretically cleaner gauge theory, where genuine nonperturbative predictions can be made and tested against numerical data~\cite{Fernandez:2022kzi,Ma:2023vgc,Albert:2023jtd,Albert:2023seb}.

In this work, we focus on the quark momentum fraction $\langle x\rangle_q$ of the pion, which measures the fraction of the pion momentum carried by quarks rather than gluons; this quantity is accessible in lattice QCD through matrix elements of the quark part of the energy-momentum tensor, see e.g. Refs.~\cite{QCDSF:2012mkm,Pefkou:2021fni,Hackett:2023nkr}. Specifically, we compute the pion momentum fraction directly at large $N_c$, providing for the first time a calculation that accounts for 
continuum and quark-mass extrapolations in the result.

Direct large-$N_c$ lattice calculations are feasible but expensive because the cost of both gauge-field generation and fermion inversions grows rapidly with $N_c$~\cite{DeGrand:2023hzz,Bali:2013kia,Baeza-Ballesteros:2022azb,Baeza-Ballesteros:2025iee}; see Ref.~\cite{Hernandez:2020tbc} for a review. As such, calculations are done at low values of $N_c$ and the limit is obtained by extrapolating. The Twisted Eguchi--Kawai (TEK) model~\cite{Eguchi:1982nm,GonzalezArroyo:1982ub,GonzalezArroyo:1982hz,GonzalezArroyo:2010ss} gives a practical way around this difficulty by using large-$N_c$ volume reduction: the gauge dynamics are represented by a one-site model with a nontrivial twist, while suitable reduced-model observables reproduce infinite-volume large-$N_c$ Yang-Mills observables. The same framework has been used to compute the meson masses, decay constants and the chiral condensate at large $N_c$~\cite{GarciaPerez:2020gnf,Bonanno:2023ypf,Bonanno:2025bla}, and has recently been extended to pion--pion scattering
\cite{Bonanno:2026gay}.

In this work, the pion quark momentum fraction is computed in the large-$N_c$ limit using pure-gauge TEK ensembles and valence maximally twisted mass fermions~\cite{Urbach:2007rt}. For the continuum extrapolation, we include several lattice spacings 
in the range ${a \sqrt{\sigma} \in [0.15, 0.25]}$, where $\sigma \sim (440  {\rm MeV})^2$ is the string tension. Moreover, we consider several pion masses varying from $m_\pi/\sqrt{\sigma} \in [1.7, 2.1]$ to perform chiral extrapolation. We also study residual finite-$N_c$ corrections within the TEK framework, and do not see a significant dependence compared to the size of the statistical uncertainty.

The relevant hadronic matrix element is obtained using the Feynman-Hellmann method on the valence fermion Dirac operator, so that the desired quantity is extracted from the response of pion two-point functions rather than from an explicit three-point-function calculation~\cite{QCDSF:2012mkm,Bouchard:2016heu,Hannaford-Gunn:2022lez,Batelaan:2023jqp}. We then convert the bare lattice result perturbatively to the $\overline{\mathrm{MS}}$ scheme at $2\,\mathrm{GeV}$. Our final large-$N_c$ result is 
\begin{equation}
    \langle x\rangle_q^{\overline{\mathrm{MS}}}(2\,\mathrm{GeV})=0.267(105),
\end{equation}
where the uncertainty includes both the weighted statistical uncertainty and the spread among the global-fit variations.

This result is significantly smaller than existing $N_c=3$ determinations at the physical pion mass. For comparison, recent $N_c=3$ determinations give $0.575(79)$ from lattice QCD~\cite{Alexandrou:2024pionMomentum} and $0.561(28)$ from the JAM global PDF analysis~\cite{Barry:2025wjx}, in the same scheme and at the same scale. While earlier calculations at finite lattice spacing suggested that the large-$N_c$ result was consistent with $N_c=3$ QCD~\cite{Karthik:2022largeNcParton}, our analysis indicates that, once lattice-spacing and other systematic effects are controlled, the near-equipartition of the pion momentum observed in QCD is likely a finite-$N_c$ phenomenon, with the pion becoming qualitatively more gluon dominated in the large-$N_c$ limit.

The remainder of this paper is organized as follows. In Sec.~\ref{sec:setup} we describe the TEK gauge ensembles, the reduced fermion construction, the Feynman-Hellmann perturbation we use, and the perturbative renormalization factor. In Sec.~\ref{sec:results} we describe the extraction of the Feynman-Hellmann slope from the perturbed pion correlators. In Sec.~\ref{sec:momfrac} we present the extrapolation to the continuum and physical pion mass at large $N_c$, and compare the result with existing information on pion structure.

\section{Lattice setup}\label{sec:setup}

\subsection{Twisted Eguchi-Kawai reduction}
The Twisted Eguchi-Kawai (TEK) model utilizes the idea of volume independence in the large-$N_c$ limit, originating with Eguchi and Kawai~\cite{Eguchi:1982nm}. Although the original one-site reduction fails in the continuum limit when the required center symmetry is broken~\cite{Bhanot:1982sh}, the twisted reduced model of Refs.~\cite{GonzalezArroyo:1982ub,GonzalezArroyo:1982hz} preserves the reduction by using appropriate twisted boundary conditions. This enables the study of large-$N_c$ lattice gauge theories on a reduced model with only a single lattice point. Volume reduction states that suitable observables (e.g.\ Wilson loops) in this reduced model reproduce the corresponding observables of the infinite-volume large-$N_c$ Yang-Mills theory as $N_c \to \infty$~\cite{GonzalezArroyo:2010ss,GarciaPerez:2020gnf}.

The gauge degrees of freedom in the TEK model consist of four $\mathrm{SU}(N_c)$ matrices $U_\mu$ (with no space-time label), with pure-gauge action given by
\begin{equation}
    S_{\rm TEK}
    =
    -b N_c
    \sum_{\mu\neq\nu}
    z_{\nu\mu}
    \Tr\left(
        U_\mu U_\nu U_\mu^\dagger U_\nu^\dagger
    \right)
\end{equation}
where $b := 1/(g^2 N_c)$ is the inverse 't Hooft coupling, and the \textit{twist phases} are roots of unity representing the phase-twisted boundary conditions in the one-site model,
\begin{equation}
    z_{\mu\nu}
    =
    \exp\left(\frac{2\pi i n_{\mu\nu}}{N_c}\right),
    \qquad
    n_{\mu\nu}=-n_{\nu\mu}\in\mathbb{Z}.
\end{equation}
The integer $n_{\mu\nu}$ tensor may be freely chosen up to certain conditions (see~\cite{Gonzalez-Arroyo:1997ugn,GarciaPerez:2020gnf}); in this work we always choose $N_c$ as a perfect square $N_c = L^2$ and use the symmetric twist 
\begin{equation}
    n_{\mu\nu} = k L
    \qquad (\mu>\nu)
    \label{eq:symmetric_twist}
\end{equation}
with $k$ as an integer coprime to $L$, with particular values chosen following the ensembles in Ref.~\cite{GarciaPerez:2020gnf} to avoid symmetry-breaking issues.

The gauge configurations are sampled according to the partition function
\begin{equation}
    Z_{\rm TEK}
    =
    \prod_{\mu=1}^{4}
    \int dU_\mu \, e^{-S_{\rm TEK}}.
\end{equation}
Notably, the quark determinant has been dropped, as we are studying QCD in the 't Hooft limit of fixed $g^2N_c$ and number of flavors $N_f$ held constant, where the theory becomes quenched to leading order in the $1/N_c$ expansion.

\subsection{Ensembles and scale setting}

The ensembles used in this work are summarized in Table~\ref{tab:ensembles}. All ensembles use the symmetric twist with flux parameter $k=7$. The matrix sizes are $N_c = 361$ and $N_c = 529$, corresponding to effective linear sizes $L=19$ and $L=23$. The gauge-field generation follows the same TEK overrelaxation strategy used in Refs.~\cite{Perez:2015ssa,GarciaPerez:2020gnf}, where the number of sweeps between stored configurations was chosen large ($\sim 5 \cdot 10^3$) enough that Monte Carlo time dependence was not visible in the measured observables. The scale is quoted through the string tension in lattice units, $a\sqrt{\sigma}$, using the string-tension determination of Ref.~\cite{GonzalezArroyo:2012fx}, as tabulated for the TEK meson-spectrum ensembles in Ref.~\cite{GarciaPerez:2020gnf}.\footnote{For a detailed discussion of scale setting using gradient
flow and Wilson loops in twisted reduced models, see
Ref.~\cite{Butti:2022sgy}.} For each gauge ensemble, the valence twisted mass is tuned to one or more of the approximate pion-mass targets $m_\pi/\sqrt{\sigma}=1.7,1.9,2.1$. As done in Ref.~\cite{GarciaPerez:2020gnf}, we set the physical scale by taking $\sqrt{\sigma}=440\text{ MeV}$, so that the pion mass targets correspond to $m_\pi \approx 748\text{ MeV}, 836\text{ MeV}, 924\text{ MeV}$ respectively.\footnote{Although in a large-$N_c$ setting the value $\sqrt{\sigma}=440\text{ MeV}$ is a conventional scale choice rather than a directly physical quantity, it is also not a very sensitive choice for our final results. For instance, a $10\%$ change in this scale results in our final result shifting by less than $1\%$ of its central value, which is well within the statistical uncertainty.}

\begin{table*}[t]
    \centering
    \caption{Gauge ensembles used in the analysis; the parameters are matched to those of Ref.~\cite{GarciaPerez:2020gnf}.
    For each ensemble, the number of sweep steps between stored configurations is $N_\mathrm{sweeps} \sim 5 \cdot 10^3$, much larger than the autocorrelation time. Valence twisted-mass parameters and fitted pion masses used for the gauge ensembles are shown in panel (b). 
    For cases where $\kappa_c$ was not taken from an existing determination, we tuned it so that $|m_{PCAC}| / \mu < 0.1$, with $m_{PCAC}$ statistically consistent with zero.}
    \label{tab:ensembles}
    \textbf{(a) Gauge ensembles}
    
    \vspace{0.3em}
    \begin{tabular}{cccccccc}
        \toprule
        $N_c$ & $L$ & $k$ & $N_{\rm therm}$ & $N_{\rm cfg}$ & $b$ & $a\sqrt{\sigma}$ & $\kappa_c$ \\
        \midrule
        361 & 19 & 7 & 20000 & 4000 & 0.355 & 0.2410(31) & 0.162978 \\
        361 & 19 & 7 & 20000 & 4000 & 0.360 & 0.2058(25) & 0.160243 \\
        361 & 19 & 7 & 20000 & 4800 & 0.365 & 0.1784(17) & 0.158008 \\
        361 & 19 & 7 & 20000 & 5600 & 0.370 & 0.1573(18) & 0.15638 \\
        529 & 23 & 7 & 25000 & 3200 & 0.365 & 0.1784(17) & 0.15796 \\
        529 & 23 & 7 & 25000 & 6400 & 0.370 & 0.1573(18) & 0.15624 \\
        \bottomrule
    \end{tabular}
    
    \vspace{1em}
    \textbf{(b) Valence twisted-mass parameters}
    
    \vspace{0.3em}
    \begin{tabular}{ccccc}
        \toprule
        $N_c$ & $b$ & $m_\pi/\sqrt{\sigma}$ targets & fitted $am_\pi$ & $\mu$ \\
        \midrule
        361 & 0.355 & $\{1.7,1.9,2.1\}$ & $\{0.3971(76),0.4583(39),0.497(19)\}$ & $\{0.036523,0.045622,0.055732\}$ \\
        361 & 0.360 & $\{1.7,1.9,2.1\}$ & $\{0.351(12),0.3919(97),0.436(11)\}$ & $\{0.031336,0.039143,0.047817\}$ \\
        361 & 0.365 & $\{1.7,1.9,2.1\}$ & $\{0.3190(47),0.3469(42),0.3914(82)\}$ & $\{0.027395,0.034220,0.041804\}$ \\
        361 & 0.370 & $\{1.7,1.9,2.1\}$ & $\{0.2804(93),0.2946(78),0.320(13)\}$ & $\{0.020531,0.025646,0.031329\}$ \\
        529 & 0.365 & $\{1.7,1.9\}$ & $\{0.2984(46),0.3360(44)\}$ & $\{0.027979,0.034950\}$ \\
        529 & 0.370 & $\{1.7,1.9,2.1\}$ & $\{0.2693(85),0.297(12),0.3385(80)\}$ & $\{0.024937,0.031150,0.038053\}$ \\
        \bottomrule
    \end{tabular}
\end{table*}

\subsection{TEK fermions and pion correlators}\label{sec:tek_fermions_pion_correlators}
In order to compute fermionic correlation functions, we introduce the four $N_c \times N_c$ unitary ``twist-eater matrices'' $\Gamma_\mu$, constructed to satisfy
\begin{equation}\label{eq:twist_eater}
    \Gamma_\mu\Gamma_\nu
    =
    z_{\mu\nu}\Gamma_\nu\Gamma_\mu
\end{equation}
These matrices are unique up to unitary conjugation and multiplication by a phase~\cite{Gonzalez-Arroyo:1997ugn,GarciaPerez:2020gnf}, and play the role of restoring an effective translation structure seen by the quark fields. In our case where $N_c=L^2$ and $n_{\mu\nu}$ is the symmetric twist~\eqref{eq:symmetric_twist}, an explicit example of a representation can be constructed from the two $L\times L$ matrices
\begin{equation}
    P_{ij}=\delta_{j,(i+1)\bmod L},
    \qquad
    Q_{ij}=\exp\left(\frac{2\pi i k i}{L}\right)\delta_{ij}
\end{equation}
as
\begin{equation}
    \begin{aligned}
    \Gamma_1&=\mathbbm{1}_L\otimes Q,
    &
    \Gamma_2&=Q\otimes QP,
    \\
    \Gamma_3&=P\otimes QP,
    &
    \Gamma_4&=\mathbbm{1}_L\otimes P.
    \end{aligned}
\end{equation}
The effective space-time dependence of the valence quark propagator is carried by these matrices as described in Refs.~\cite{Gonzalez-Arroyo:2015bya,GarciaPerez:2020gnf}, allowing us to compute spacetime-separated fermionic correlators despite working on a single-site lattice. The TEK Wilson-Dirac operator is given in momentum space as an operator on the tensor product of the spin, color, and twist-eater spaces:
\begin{align}
    D_W(p)
    &=
    \mathbbm{1}
    -
    \kappa
    \sum_{\mu=1}^{4}
    \bigg[
        e^{iap_\mu}
        (\mathbbm{1}-\gamma_\mu)
        \otimes U_\mu\otimes\Gamma_\mu^*
        \nonumber\\
    &\hspace{1.5cm}
        +
        e^{-iap_\mu}
        (\mathbbm{1}+\gamma_\mu)
        \otimes U_\mu^\dagger\otimes\Gamma_\mu^t
    \bigg].
    \label{eq:DW_TEK}
\end{align}
The factors $U_\mu\otimes\Gamma_\mu^*$ and $U_\mu^\dagger\otimes\Gamma_\mu^t$ are essentially the reduced-model replacements for the forward and backward parallel transporters in the ordinary Wilson operator. The momentum $p_\mu$ lives on a lattice corresponding to the finite-volume box that the quarks live in. In this paper, we work with quark bilinear operators in the pion channel projected to zero spatial momentum, and temporal momentum $p_4$ living on a lattice with effective temporal size $2L$:
\begin{equation}
    p_4=\frac{\pi n_4}{aL},
    \qquad
    n_4=0,\ldots,2L-1.
\end{equation}
The pion correlator for temporal separation $t$ is then given by~\cite{Gonzalez-Arroyo:2015bya} (unlike this previous work, we use unsmeared correlators)
\begin{align}
    C_\pi(t)
    &=
    \frac{1}{2L}
    \sum_{n_4=0}^{2L-1}
    e^{-(2 \pi i ) t n_4/ (2L)}
    \nonumber\\
    &\hspace{0.5cm}\times
    \left\langle
        \Tr\left[
            \gamma_5
            \left(D_W(0)\right)^{-1}
            \gamma_5
            \left(D_W(p_4)\right)^{-1}
        \right]
    \right\rangle,
    \label{eq:TEK_pion_corr}
\end{align}
where the trace is over spin, color, and twist-eater indices. As in Ref.~\cite{GarciaPerez:2020gnf}, this trace is estimated using $Z_4$ random sources, which replace the trace by an average
\begin{equation}
    \mathbb{E}_{z} \left[ z^\dagger \gamma_5 D_W(0)^{-1} \gamma_5 D_W(p_4)^{-1} z \right]
\end{equation}
across a stochastic source $z$ living in $\text{spin}\otimes\text{color}\otimes\text{twist}$ space with entries drawn uniformly as $z_{\alpha \beta \gamma} \in \{ \frac{1}{\sqrt{2}} (\pm 1 \pm i) \}$.
The reduced-model correlator also involves an internal quark-momentum average; again following Ref.~\cite{GarciaPerez:2020gnf}, this is simulated by multiplying the TEK links by random $U(1)$ phases,
\begin{equation}
    U_\mu \to e^{i\alpha_\mu}U_\mu,
\end{equation}
with the angles $\alpha_\mu$ chosen uniformly for each $\mu$. This provides a stochastic estimate of the momentum sum at no additional inversion cost.

Finally, in this work we use maximally twisted mass fermions, instead of ordinary Wilson fermions, as their scaling towards the continuum limit is $O(a^2)$~\cite{Frezzotti:2003ni}.

The sign of the twisted mass in the Dirac operator is different for each of the light quarks,
\begin{equation}
    D^\pm(p)
    =
    D_W(p)\big|_{\kappa=\kappa_c}
    \pm i\,2\kappa_c\mu \gamma_5 .
    \label{eq:DTM_TEK}
\end{equation}
The twisted-mass version of the trace inside Eq.~\eqref{eq:TEK_pion_corr} is
\begin{equation}
    \begin{aligned}
    &\Tr\left[
        \gamma_5
        \left(D^-(0)\right)^{-1}
        \gamma_5
        \left(D^+(p_4)\right)^{-1}
    \right]
    \\
    &=
    \mathbb{E}_{z} \left[
        z^\dagger \gamma_5 \left(D^-(0)\right)^{-1}
        \gamma_5 \left(D^+(p_4)\right)^{-1} z
    \right].
    \end{aligned}
\end{equation}
We compute this in practice as $\mathbb{E}_{z} \left[ x(0)^\dagger x(p_4) \right]$ where $x(p_4) := \gamma_5 \left(D^+(p_4)\right)^{-1} z$ is obtained via conjugate gradient inversion.

\subsection{Feynman-Hellmann construction for the quark momentum fraction}

A Feynman-Hellmann approach is used to extract matrix elements of the quark part of the energy-momentum tensor. As discussed in Ref.~\cite{QCDSF:2012mkm}, for any perturbation $S(\delta)=S+\delta \cdot S_O$ of the action $S$ parametrized by a small parameter $\delta$, the Feynman-Hellmann relation gives
\begin{equation}\label{eq:FH_general}
    \left.\frac{\partial m_H(\delta)}{\partial \delta}\right|_{\delta=0}
    =
    \frac{1}{2m_H}
    \bra{H(\vec{p}=0)}: O:\ket{H(\vec{p}=0)} ,
\end{equation}
where $O$ is the local operator whose spacetime sum gives $S_O$.
Here $H$ is any hadron state, which is taken below to be the pion, $m_H(\delta)$ is the ground-state energy of that hadron under the perturbed action, and $:O:$ is the normal-ordered version of the operator appearing in the perturbation. For the present matrix element the perturbing operator is
\begin{equation}\label{eq:SO_def}
    \begin{aligned}
        O
        &:=
        T^q_{00}-\frac13\sum_i T^q_{ii}
        \\
        &=
        \bar{\psi}
        \left(
            \gamma_0\overleftrightarrow{D}_0
            -\frac13\sum_i\gamma_i\overleftrightarrow{D}_i
        \right)
        \psi ,
    \end{aligned}
\end{equation}
where
\begin{equation}
        \overleftrightarrow{D}_{\mu}
        =
        \frac{\overrightarrow{D}_\mu-\overleftarrow{D}_\mu}{2}.
\end{equation}
The operator $O$ has vanishing vacuum expectation value by symmetry. The right hand side of Eq.~\eqref{eq:FH_general} is therefore
\begin{equation}
    \frac{1}{2m_\pi}
    \bra{\pi(\vec{p}=0)}:O:\ket{\pi(\vec{p}=0)}
    =
    -m_\pi \langle x\rangle_q,
\end{equation}
using the standard expression
\begin{equation}
    \bra{\pi(p)} T_{\mu\mu}^q \ket{\pi(p)}
    =
    2\langle x\rangle_q^{\mathrm{lat}}
    \left(p_\mu p_\mu-\frac{p^2}{4}\right).
\end{equation}
Thus, for the bare lattice quantity,
\begin{equation}\label{eq:FH_mpi_deriv}
    \langle x\rangle_q^{\mathrm{lat}}
    =
    -\frac{1}{m_\pi}
    \left.\frac{\partial m_\pi(\delta)}{\partial\delta}\right|_{\delta=0}.
\end{equation}
This relation is used to extract $\langle x\rangle_q^{\mathrm{lat}}$ from $m_\pi(\delta)$ computed at several small values of $\delta$. In Appendix~\ref{app:free_theory}, we verify this relationship in the free theory, where the gauge fields are non-dynamical and we must have $\langle x\rangle_q^{\mathrm{lat}} = 1$ in the continuum limit.

Operationally, $m_\pi(\delta)$ is obtained by adding the corresponding lattice perturbation to the fermion action and recomputing the pion correlator. In the large-$N_c$ limit, effects from the fermion determinant, accounting for disconnected diagrams, are subleading. Before reduction, the perturbation has the hopping-term form
\begin{equation}
    \delta S_O
    =
    \frac12
    \sum_x\sum_{\mu=\pm 1}^{\pm 4}
    \bar{\psi}(x)\delta_\mu\gamma_\mu U_\mu(x)\psi(x+\hat{\mu}),
\end{equation}
where $\delta_{\pm 4}=\delta$, $\delta_{\pm i}=-\delta/3$ for spatial directions $i=1,2,3$, and $\gamma_{-\mu}:=-\gamma_\mu$. With the usual Wilson normalization of the fermion fields in terms of the hopping parameter $\kappa$, the full fermionic part of the action (of a non-volume-reduced model) is
\begin{equation}
    \begin{aligned}
        S_f
        &=
        \sum_x
        \left[
            \bar{\psi}(x)\psi(x)
        \right.
        \\
        &\hspace{1.5cm}
        \left.
            -
            \kappa\sum_{\mu=\pm 1}^{\pm 4}
            \bar{\psi}(x)
            (\mathbbm{1}-\gamma_\mu)
            U_\mu(x)\psi(x+\hat{\mu})
        \right].
    \end{aligned}
\end{equation}
The effect is therefore to replace the $\gamma_\mu$ factors in $S_f$ by $(1+\delta_\mu)\gamma_\mu$. Applying the same reduced-model replacement as in Eq.~\eqref{eq:DW_TEK} gives the perturbed TEK Wilson-Dirac operator
\begin{align}
    D_W(p,\delta)
    &=
    \mathbbm{1}
    -
    \kappa
    \sum_{\mu=1}^{4}
    \bigg[
        \nonumber\\
    &\hspace{0.5cm}
        e^{iap_\mu}
        \big(\mathbbm{1}-(1+\delta_\mu)\gamma_\mu\big)
        \otimes U_\mu\otimes\Gamma_\mu^*
        \nonumber\\
    &\hspace{0.5cm}
        +
        e^{-iap_\mu}
        \big(\mathbbm{1}+(1+\delta_\mu)\gamma_\mu\big)
        \otimes U_\mu^\dagger\otimes\Gamma_\mu^t
    \bigg].
    \label{eq:DW_TEK_FH}
\end{align}
As previously mentioned in subsection~\ref{sec:tek_fermions_pion_correlators}, in this work we use valence fermions with maximal twist, so that we are working with the operators
\begin{equation}
    D^\pm(p,\delta)
    =
    D_W(p,\delta)\big|_{\kappa=\kappa_c}
    \pm i\,2\kappa_c\mu \gamma_5 .
\end{equation}
This perturbed twisted-mass Dirac operator is used to compute the pion correlators, from which the $m_\pi(\delta)$ values and hence $\langle x\rangle_q^{\mathrm{lat}}$ are extracted.

\subsection{Perturbative renormalization of $\langle x\rangle_q$}

The Feynman-Hellmann construction described above gives the bare lattice quark momentum fraction. We apply a perturbative renormalization factor to convert this to the $\overline{\mathrm{MS}}$ scheme at $2\,\mathrm{GeV}$.\footnote{Note that TEK is (in the large-$N_c$ limit) a volume-reduced formulation of the same lattice gauge theory, and in particular, has the same discretization behavior. Hence we may take our renormalization factors from ordinary unreduced large-$N_c$ Wilson QCD.}
In quenched or large-$N_c$ QCD, one has
\begin{align}
    \langle x\rangle_q^{\overline{\mathrm{MS}}}(M)
    &=
    \left(Z_{\mathrm{lat}}^{\overline{\mathrm{MS}}}\right)_{qq}(M)
    \langle x\rangle_q^{\mathrm{bare}}
\end{align}
with no additional mixing from the gluonic part~\cite{QCDSF:2012mkm}. The perturbative renormalization factor of the quark-momentum fraction operator we use is~\cite{Gockeler:2004wp},
 \begin{equation}
\begin{aligned}
&Z_{\mathrm{lat}}^{\msbar}(M,a)
  \stackrel{N_c \to \infty}{\longrightarrow} {}\\
&\quad 1 - \frac{1}{2(4\pi)^2 b}
  \left[\frac{16}{3}\ln(aM) + 2.56184\right]
  + O(g_0^4).
\end{aligned}
\end{equation}
This expression corresponds to the operator denoted as $\mathcal{O}_{v_{2b}}$ in Ref.~\cite{Gockeler:2004wp}, using $c_{sw}=0$ and $N_c \to \infty$.

\section{Correlator analysis}\label{sec:results}

For each ensemble and valence pion mass, we compute pion correlators at the small perturbation values $\delta = -\lambda, 0, \lambda$ with $\lambda = 0.01$. 

At large time separations, the correlator is dominated by the ground-state contribution, which (in addition to a constant background, see~\cite{GarciaPerez:2020gnf})
takes the form
\begin{equation}
    C_\delta(t)
    \sim
    A_\delta
    \cosh\left[
        m_\delta \left(t - \frac{T}{2}\right)
    \right]
    +
    B_\delta.
\end{equation}
To begin the procedure of extracting ground-state masses from these correlators, we remove the additive constant background by working with the temporal finite difference
\begin{equation}
    \Delta C_\delta(t)
    :=
    C_\delta(t)-C_\delta(t+1)
    \label{eq:deltaC_def}
\end{equation}
which removes $B_\delta$ and has large-time asymptotic form
\begin{equation}
    \Delta C_\delta(t)
    \sim
    2A_\delta F(m_\delta,t),
    \label{eq:single_state_deltaC}
\end{equation}
where for convenience we have defined
\begin{equation}
    F(m,t)
    =
    \sinh\left(\frac{m}{2}\right)
    \sinh\left[
        \frac{m}{2}(T-2t-1)
    \right].
    \label{eq:F_shape}
\end{equation}
Similarly, we can have a two-state fit to $\Delta C_\delta(t)$ by including an additional term,
\begin{equation}
    \Delta C_\delta(t)
    \sim
    2A_{0\delta} F(m_{0\delta},t)
    +
    2A_{1\delta} F(m_{1\delta},t).
    \label{eq:two_state_deltaC}
\end{equation}
We use this two-state fit to extract the unperturbed meson ground state mass $m_{0,\delta=0}$ by fitting $\Delta C_0(t)$ over several fit windows $t\in[t_{\min},t_{\max}]$, with $t_{\max}$ held fixed near the maximum time separation, and $t_{\min}$ varied to find a plateau of stability.

Next we describe how we numerically obtain the Feynman-Hellmann derivative $\left.\frac{\partial m_\pi(\delta)}{\partial\delta}\right|_{\delta=0}$. We begin by constructing the ratio
\begin{equation}
    R(t)
    =
    \frac{
        \Delta C_+(t)-\Delta C_-(t)
    }{
        2\lambda \,\Delta C_0(t)
    } \stackrel{\lambda \to 0}{\sim} \partial_\delta\log\Delta C_\delta(t)\big|_{\delta=0}.
    \label{eq:R_ratio_def}
\end{equation}
As a check of the finite-difference step size, we also repeated the extraction using the one-sided estimator
\begin{equation}
    R_+(t)
    =
    \frac{
        \Delta C_+(t)-\Delta C_0(t)
    }{
        \lambda \,\Delta C_0(t)
    },
\end{equation}
which differs from Eq.~\eqref{eq:R_ratio_def} by terms of order $\lambda$. The resulting individual momentum fractions deviate from the symmetric-difference values by less than $0.015$; as this $O(\lambda)$ difference is smaller than the statistical uncertainty in our final results, we consider the symmetric-difference $O(\lambda^2)$ estimator to be sufficient for our purposes.
Taking the $\delta$ derivative of Eq.~\eqref{eq:single_state_deltaC} gives the single-state asymptotic for $R(t)$,
\begin{equation}
\begin{gathered}
    R(t)
    \sim
    \frac{A'_0}{A_0}
    +
    m'_0 \left.\partial_m\log F(m,t)\right|_{m=m_0}.
\end{gathered}
\label{eq:single_state_R_fit}
\end{equation}
When fitting $R(t)$ to this form, we fix $m_0$ to the value $m_{0,\delta=0}$ obtained from the previous two-state fit to $\Delta C_0(t)$ (Eq.~\eqref{eq:two_state_deltaC}), so that there are only two free parameters $A'_0/A_0$ and $m'_0$. The resulting $m'_0$ is exactly the desired Feynman-Hellmann derivative, and $-m'_0 / m_{0,\delta=0}$ gives the bare lattice quark momentum fraction.

\begin{figure*}
    \centering
    \vspace{1cm}    \includegraphics[width=\textwidth]{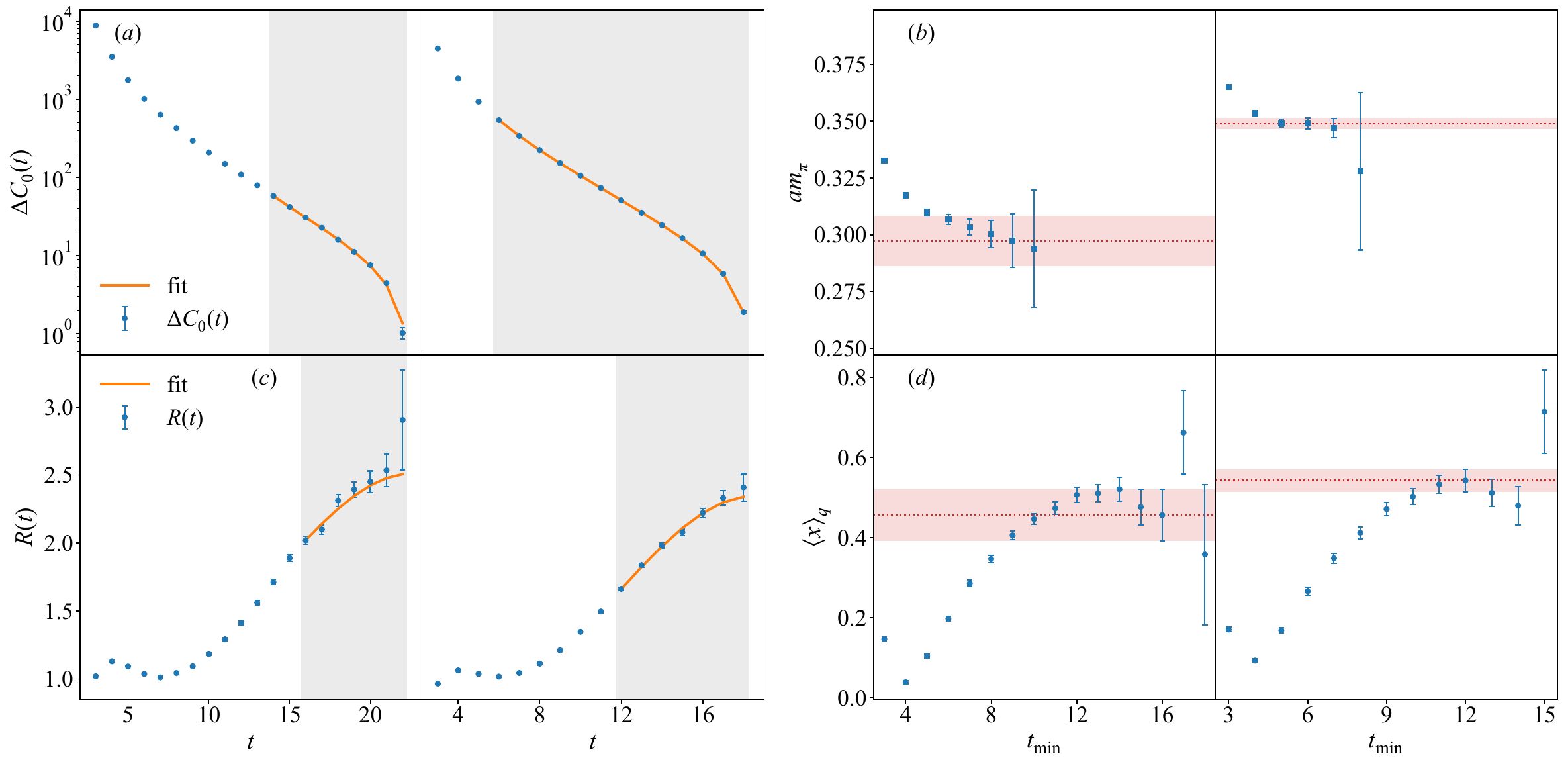}
    \vspace{0.1cm}
    \caption{
        Representative correlator-analysis fits for two ensembles, both with
        $m_\pi/\sqrt{\sigma}=1.9$. Within each adjacent pair, the left panel
        shows the ensemble with $N_c=529$ and $b=0.370$, while the right panel
        shows the ensemble with $N_c=361$ and $b=0.365$. Panel (a) shows the unperturbed
        finite-difference correlator $\Delta C_0(t)$ with a two-state fit. Panel (b) shows the fitted
        ground-state mass as the left endpoint of the two-state mass-fit window
        is varied. Panel (c) shows $R(t)$ (the response of $\log \Delta C(t)$ to the Feynman--Hellmann perturbation) and the fit used to extract the
        Feynman--Hellmann slope $m'_0$, and panel (d) shows the renormalized momentum fraction as the
        left endpoint of the Feynman--Hellmann fit window is varied. 
        The grey shaded regions in panels (a) and (c) indicate the selected fit windows; the horizontal bands in panels (b) and (d) show the selected-window values and their statistical uncertainties.
    }
    \label{fig:correlator_fit_results}
\end{figure*}

\section{The pion momentum fraction}\label{sec:momfrac}

\subsection{Global fit extrapolation}
\begin{figure}[h!]
    \centering
    \vspace{0.85cm}    \includegraphics[width=\columnwidth]{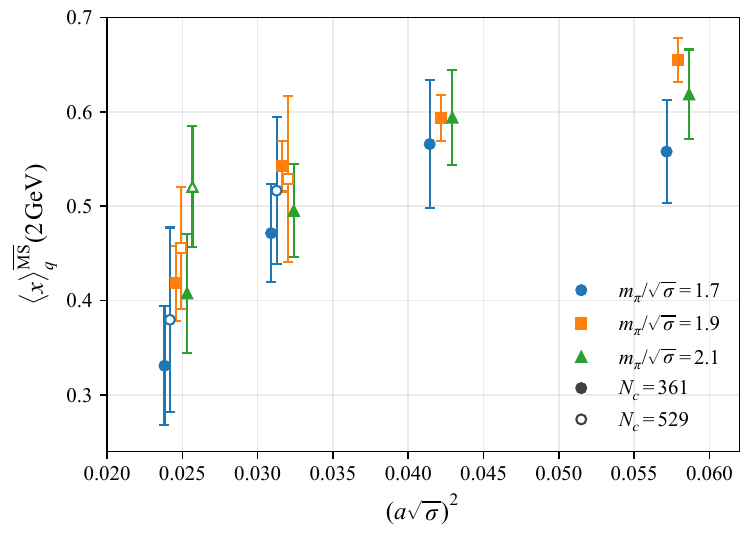}
    \caption{Renormalized momentum-fraction values entering the global fit. Points at the same lattice spacing are shifted slightly horizontally for visibility.} 
    \label{fig:qmf_input_points}
\end{figure}

We now combine the Feynman-Hellmann slopes extracted in Sec.~\ref{sec:results} to obtain the pion quark momentum fraction. For each parameter combination $(m_\pi/\sqrt{\sigma},N_c, b)$, the Feynman-Hellmann slope is taken from a fit range chosen from the corresponding stability plots. The quantity entering the global analysis is the renormalized momentum fraction
\begin{align}
    \langle x\rangle_{q,i}^{\overline{\mathrm{MS}}}(2\,\mathrm{GeV})
    = 
        -\frac{1}{m_{\pi,i}}
        \left.
        \frac{\partial m_{\pi,i}}{\partial\delta}
        \right|_{\delta=0}
    Z_{\mathrm{lat}}^{\overline{\mathrm{MS}}}
    (2\,\mathrm{GeV},a_i) , 
    \label{eq:renorm_qmf_global}
\end{align}
where the renormalization factor is the perturbative factor described above. The statistical covariance used in the extrapolation is obtained from the same bootstrap samples used for the Feynman-Hellmann extraction. The resulting values are listed in Table~\ref{tab:qmf_inputs} for a particular chosen collection of $t_{\min}^{R}$; we use all these values in a baseline global fit, hereafter referred to as the ``central" fit.
The same data is also shown in Fig.~\ref{fig:qmf_input_points}. As can be seen, the main trend in the data seems to be the lattice spacing dependence, while the pion mass dependence seems milder. As for the $N_c$ dependence, it can be seen that points with $N_c=361$ and $N_c=529$ are statistically consistent within $\approx 1\sigma$.

\begin{table}[h!]
    \centering
    \vspace{0.2cm}
    \caption{Renormalized momentum-fraction values entering the central global fit. The quoted uncertainties are statistical bootstrap errors from the Feynman-Hellmann extraction.}
    \label{tab:qmf_inputs}
    \setlength{\tabcolsep}{3pt}
    \begin{tabular}{ccccc}
        \toprule
        target $m_\pi/\sqrt{\sigma}$ & $N_c$ & $b$ & $t_{\min}^{R}$ & $\langle x\rangle_{q,i}^{\overline{\mathrm{MS}}}(2\,\mathrm{GeV})$ \\
        \midrule
        1.7 & 361 & 0.355 & 12 & $0.558(55)$ \\
        1.7 & 361 & 0.360 & 12 & $0.566(68)$ \\
        1.7 & 361 & 0.365 & 12 & $0.471(52)$ \\
        1.7 & 361 & 0.370 & 12 & $0.331(63)$ \\
        1.7 & 529 & 0.365 & 16 & $0.517(78)$ \\
        1.7 & 529 & 0.370 & 16 & $0.380(98)$ \\
        1.9 & 361 & 0.355 & 12 & $0.655(23)$ \\
        1.9 & 361 & 0.360 & 12 & $0.594(24)$ \\
        1.9 & 361 & 0.365 & 12 & $0.542(27)$ \\
        1.9 & 361 & 0.370 & 12 & $0.418(39)$ \\
        1.9 & 529 & 0.365 & 16 & $0.529(88)$ \\
        1.9 & 529 & 0.370 & 16 & $0.456(65)$ \\
        2.1 & 361 & 0.355 & 11 & $0.618(48)$ \\
        2.1 & 361 & 0.360 & 11 & $0.594(51)$ \\
        2.1 & 361 & 0.365 & 11 & $0.495(49)$ \\
        2.1 & 361 & 0.370 & 11 & $0.408(63)$ \\
        2.1 & 529 & 0.370 & 15 & $0.521(64)$ \\
        \bottomrule
    \end{tabular}
\end{table}

The chiral-continuum extrapolation is performed via a correlated global fit linear in $(a\sqrt{\sigma})^2$ and $(m_\pi/\sqrt{\sigma})^2$,
\begin{equation}
    \langle x \rangle(a,m_\pi)
    =
    A
    +
    B(a\sqrt{\sigma})^2
    +
    C\left(\frac{m_\pi}{\sqrt{\sigma}}\right)^2 .
    \label{eq:qmf_global_fit}
\end{equation}
The fit is evaluated at $a=0$ and at the physical pion mass, using the same scale setting through $\sqrt{\sigma}$ as in the rest of the analysis. The central fit from the values in Table~\ref{tab:qmf_inputs} gives ${\langle x\rangle_q^{\overline{\mathrm{MS}}}(2\,\mathrm{GeV})=0.226(66)}$, with $\chi^2/\mathrm{dof}=12.77/14$.

\begin{table*}[t]
    \centering
    \caption{
        Central fit and fit variations to account for systematic uncertainties in the extrapolation. For each fit, we define $\mathrm{AIC}_i=\chi_i^2-2\,\mathrm{dof}_i$, and the weights
        $w_i\propto\exp(-\mathrm{AIC}_i/2)$. This definition accounts for both the
        number of fit parameters and the number $N_{\rm cut}$ of removed $(N_c,b,m_\pi)$
         simulation points. From coarsest to finest, the lattice spacings correspond to inverse gauge couplings $b = 0.355$, $0.360$, $0.365$, and $0.370$.
    }
    \label{tab:qmf_subset_aic}
    \setlength{\tabcolsep}{5pt}
    \begin{tabular}{@{}lcccc@{}}
        \toprule
        Global-fit variation
        & Result
        & \(N_{\rm cut}\)
        & \(\chi^2/\mathrm{dof}\)
        & \(w_i\) \\
        \midrule
        Central fit
        & \(0.226(66)\) & 0 & \(12.77/14\) & \(0.2082\) \\
        Remove coarsest
        & \(0.135(89)\) & 3 & \(7.79/11\) & \(0.1250\) \\
        Remove finest
        & \(0.326(77)\) & 6 & \(4.04/8\) & \(0.0406\) \\
        Remove $N_c=361$ finest
        & \(0.303(74)\) & 3 & \(5.64/11\) & \(0.3664\) \\
        Remove $N_c=529$ finest
        & \(0.235(68)\) & 3 & \(11.14/11\) & \(0.0234\) \\
        Remove $N_c=361$ two finest
        & \(0.303(97)\) & 6 & \(3.86/8\) & \(0.0444\) \\
        Include $1/N_c$ term [Eq.~\eqref{eq:qmf_global_fit_nc}]
        & \(0.309(148)\) & 0 & \(12.38/13\) & \(0.0931\) \\
        Later FH \(t_{\min}\) choice (\(t_{\min}=13\) for \(N_c=361\), \(17\) for \(N_c=529\))
        & \(0.319(77)\) & 0 & \(14.26/14\) & \(0.0989\) \\
        \midrule
        AIC average
        & \(0.267(105)\)
        & & & \\
        \bottomrule
    \end{tabular}
    \vspace{-0.15cm}
\end{table*}

To assess the systematic uncertainty associated with the extrapolation, we consider three types of variation around the central fit. First, we exclude either the coarsest lattice spacing or the smallest physical volume to probe residual discretization or $1/N_c$ effects, respectively. Second, we vary the selection of $t_{\rm min}$ values used to compute the momentum fraction on each ensemble, and repeat the central fit to assess possible excited-state contamination. Third, we adopt an alternative fit form that includes an explicit $1/N_c$ correction,
\begin{equation}
    \langle x \rangle(a,m_\pi,N_c)
    =
    A
    +
    B(a\sqrt{\sigma})^2
    +
    C\left(\frac{m_\pi}{\sqrt{\sigma}}\right)^2
    +
    \frac{D}{N_c}.
    \label{eq:qmf_global_fit_nc}
\end{equation}
Together, these fit variations allow us to incorporate systematic uncertainties associated with discretization effects, finite-$N_c$ corrections, and excited-state contamination into the final uncertainty.

\begin{figure*}[t]
    \centering
    \vspace{0.7cm}
    \includegraphics[width=\textwidth]{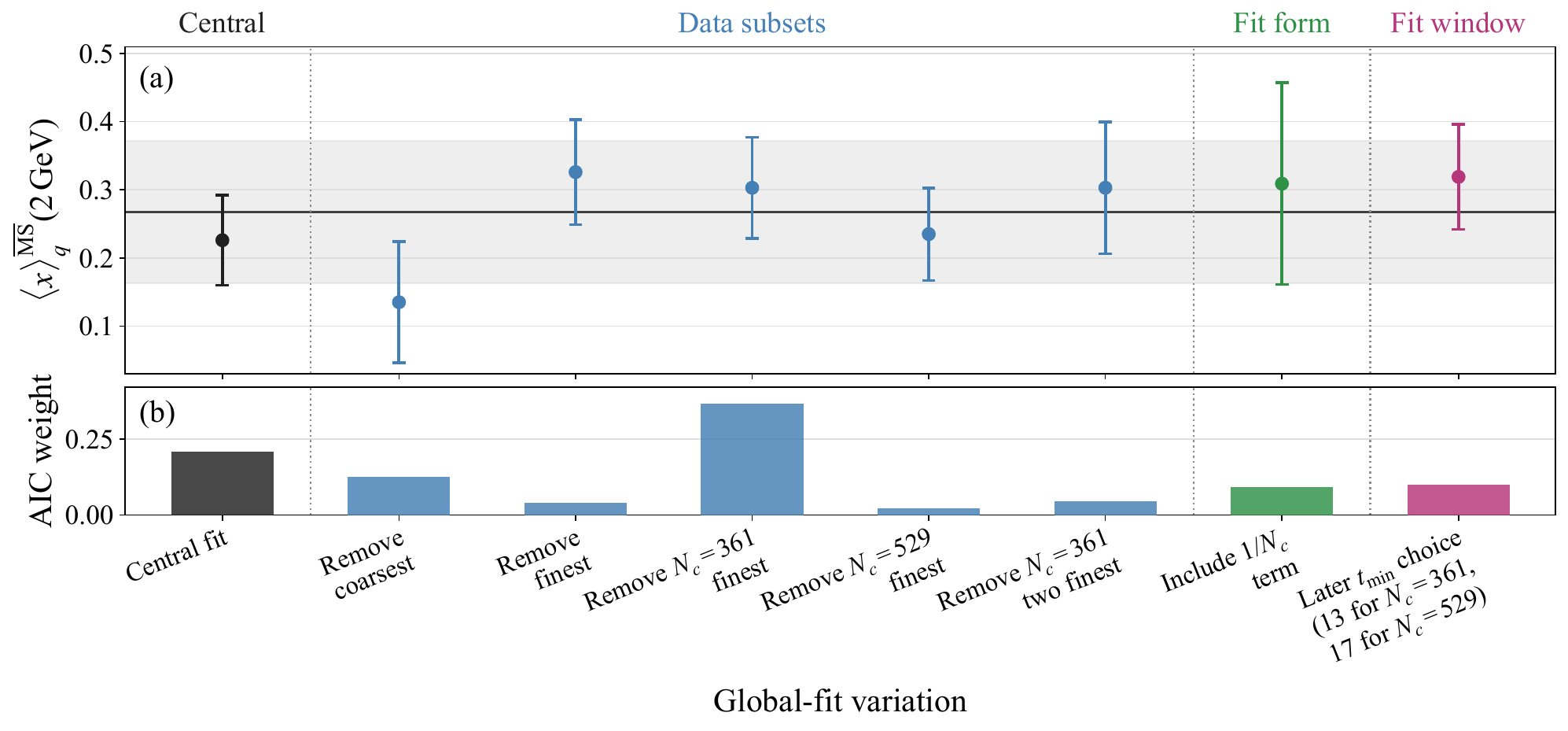}
    \caption{
        Fit results and corresponding AIC weights for the data-subset, fit-window, and fit-form variations listed in Table~\ref{tab:qmf_subset_aic}. The horizontal line and grey band indicate the AIC-averaged result and its total uncertainty, respectively.
    }
    \label{fig:qmf_ensemble_removal_aic_weights}
\end{figure*}

All fit variations are listed in Table~\ref{tab:qmf_subset_aic}. We combine their results through model averaging, with AIC weights~\cite{Jay:2020jkz}, obtaining
\begin{equation}
    \langle x\rangle_q^{\overline{\mathrm{MS}}}(2\,\mathrm{GeV})
    =
    0.267(105).
    \label{eq:qmf_final_global}
\end{equation}
The quoted uncertainty accounts for both the statistical uncertainties of the individual fits and the spread among the fit variations, weighted according to the AIC. The different results for the fit variations are represented in Fig.~\ref{fig:qmf_ensemble_removal_aic_weights}. As can be seen by the magnitude of the weights, many fit variations contribute to the weighted average.

\begin{figure*}[t]
    \centering
    \subfloat[]{
        \includegraphics[width=0.48\textwidth]{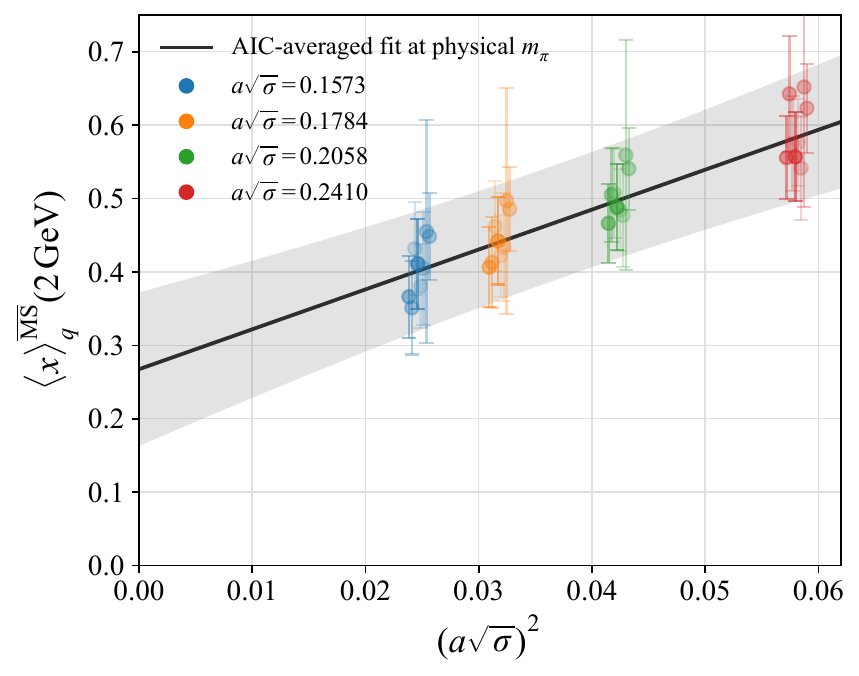}
    }
    \hfill
    \subfloat[]{
        \includegraphics[width=0.48\textwidth]{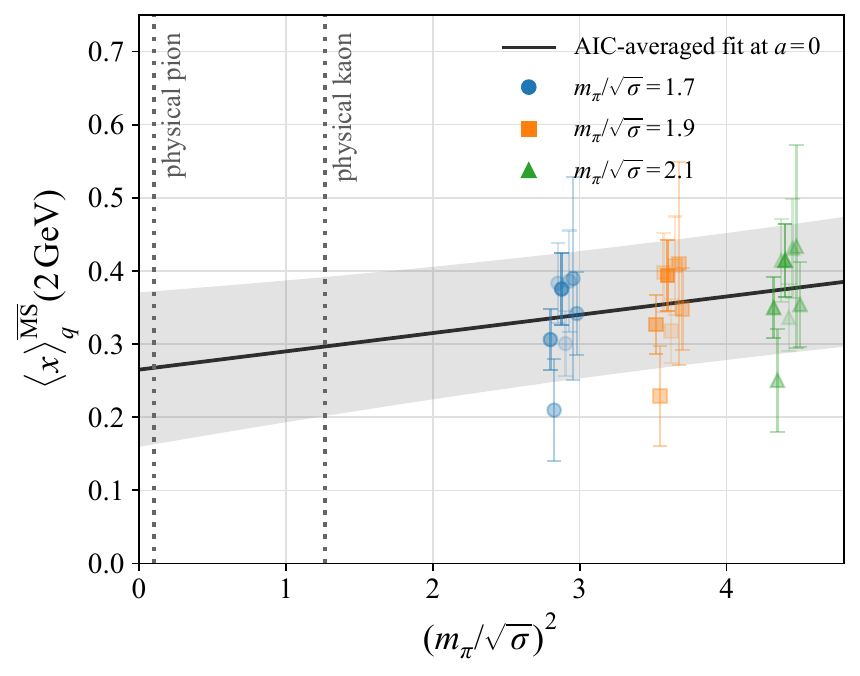}
    }
    \caption{
        Panel (a) shows the per-model predictions at the physical pion mass
        for each lattice spacing, together with the AIC-averaged global-fit
        projection at the physical pion mass. Panel (b) shows the corresponding
        per-model continuum-limit predictions at each simulated pion mass,
        together with the AIC-averaged global-fit projection at $a=0$.
        The averages use the model weights in
        Table~\ref{tab:qmf_subset_aic}; the shaded bands include
        both the weighted within-fit statistical uncertainty and the spread
        among the models. The dotted vertical lines in panel (b) mark the
        physical pion and kaon masses. The opacity of the points indicates the AIC-weight, and small horizontal offsets are applied for visibility.
    }
    \label{fig:qmf_continuum_extrapolation}
\end{figure*}

We illustrate the fit trends through two complementary projections. First, we fix the pion mass to its physical value to display the continuum extrapolation. Second, we take the continuum limit and display the pion-mass dependence. Both are shown in Fig.~\ref{fig:qmf_continuum_extrapolation} for the AIC-averaged result and the individual fit variations. The dominant extrapolation is to the continuum limit: at fixed pion mass, the renormalized momentum fraction decreases as the lattice spacing approaches zero. The pion-mass dependence over the simulated range is milder and statistically consistent with zero.

\subsection{Comparison to literature}

A direct comparison at large $N_c$ is provided by the fixed-ensemble lattice calculation of Ref.~\cite{Karthik:2022largeNcParton} which studied the pion quark structure at $N_c=17$, using Eguchi--Kawai-type volume reduction on an $8^4$ lattice as opposed to our single-site approach. Their simulation was performed at a single lattice spacing and a single pion mass, with $b=0.355$, $a\sqrt{\sigma}=0.254(2)$, and $m_\pi=0.86\sqrt{\sigma}$, resulting in a value of
\begin{equation}
    \langle x\rangle_q^{\overline{\mathrm{MS}}}(2\,\mathrm{GeV};
    a\sqrt{\sigma}=0.254,
    m_\pi/\sqrt{\sigma}=0.86)
    =
    0.50(2).
\end{equation}
Evaluating the AIC average over the global-fit variations in Table~\ref{tab:qmf_subset_aic} at the same lattice spacing and pion mass gives
\begin{equation}
    \langle x\rangle_q^{\overline{\mathrm{MS}}}
    (2\,\mathrm{GeV};
    a\sqrt{\sigma}=0.254,
    m_\pi/\sqrt{\sigma}=0.86)
    =
    0.634(87).
\end{equation}
This is more similar to the value of Ref.~\cite{Karthik:2022largeNcParton} but still higher by $\sim 1.5 \sigma$; this may be explained by several systematic differences between the two calculations, including differing fermion discretization errors (Wilson vs twisted-mass fermions) and finite-$N_c$ effects (as their calculation is not in the TEK model, and at a much smaller $N_c$).
Meanwhile, the AIC-averaged continuum extrapolation at the same pion mass yields
\begin{equation}\label{eq:qmf_continuum_at_fixed_mpi}
    \langle x \rangle (a=0,m_\pi/\sqrt{\sigma}=0.86)=0.283(99).
\end{equation}
The lower values in Eqs.~\eqref{eq:qmf_final_global} and \eqref{eq:qmf_continuum_at_fixed_mpi}
reflect a sizable downward continuum trend visible in Fig.~\ref{fig:qmf_continuum_extrapolation}, which was not reflected in the single-ensemble result of Ref.~\cite{Karthik:2022largeNcParton}.

\begin{figure*}[t]
    \centering
    \includegraphics[width=0.7\textwidth]{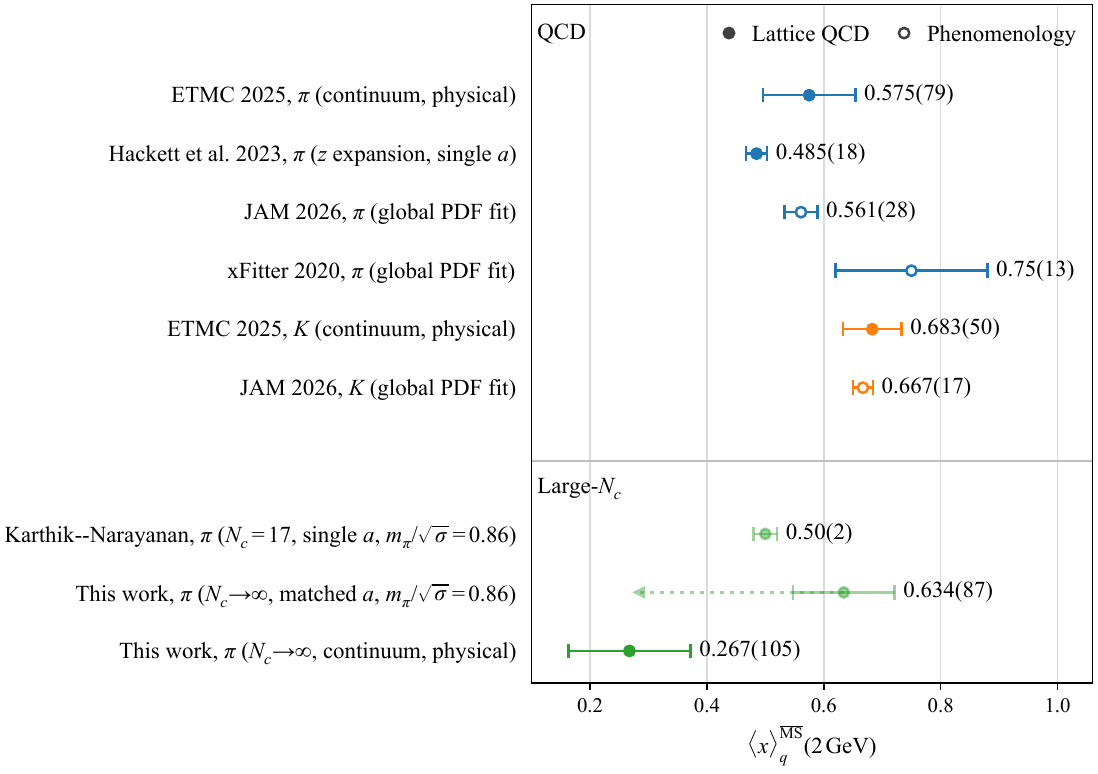}
    \caption{
        Comparison of various results for pion and kaon quark momentum fractions in
        the $\overline{\mathrm{MS}}$ scheme at $2\,\mathrm{GeV}$. The QCD
        results are taken from
        Refs.~\cite{Alexandrou:2024pionMomentum,Hackett:2023nkr,Barry:2025wjx,Novikov:2020snp}, the $N_c=17$ point is from
        Ref.~\cite{Karthik:2022largeNcParton} at $b=0.355$ and $m_\pi/\sqrt{\sigma}=0.86$, and the remaining large-$N_c$ points are AIC-averaged TEK results using the global-fit variations in Table~\ref{tab:qmf_subset_aic}.
        Solid markers denote lattice determinations, while hollow markers denote phenomenological global PDF analyses.
        In addition, we also display QCD results for the kaon quark momentum fraction from Refs.~\cite{Alexandrou:2024pionMomentum,Barry:2025wjx}. 
        The dotted green arrow represents the effect of continuum extrapolation of our matched result at fixed $m_\pi/\sqrt{\sigma}=0.86$, from $0.634(87)$ to $0.283(99)$.
    }
    \label{fig:qmf_nc_comparison}
\end{figure*}

Figure~\ref{fig:qmf_nc_comparison} displays a comparison of our large-$N_c$ results with this $N_c=17$ result, as well as various $N_c=3$ results from the literature~\cite{Alexandrou:2024pionMomentum,Hackett:2023nkr,Barry:2025wjx,Novikov:2020snp}.
Our large-$N_c$ result is significantly smaller than the continuum physical-pion $N_c=3$ result, as well as the $N_c=17$ result of Ref.~\cite{Karthik:2022largeNcParton} as discussed previously. This indicates a decrease in the pion quark momentum fraction in the large-$N_c$ limit. 

\begin{figure*}[t]
    \centering
    \includegraphics[width=0.72\textwidth]{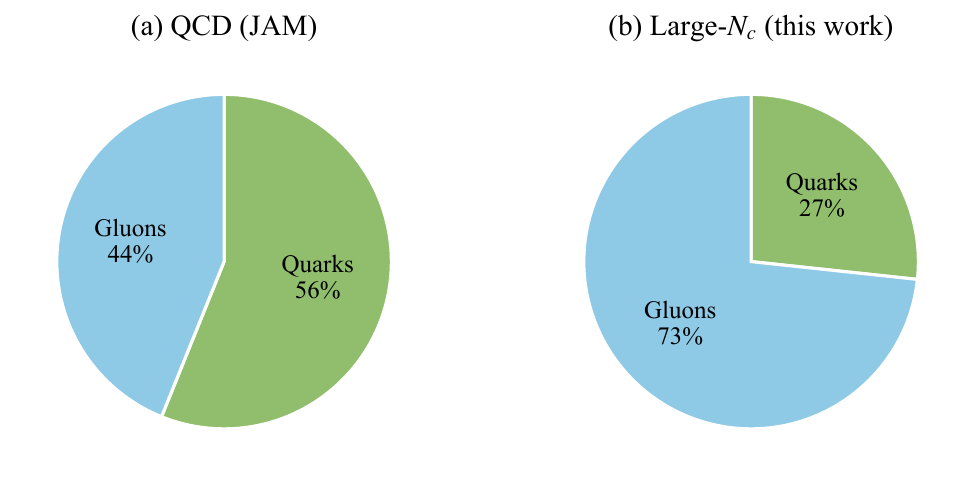}
    \caption{
        Visual breakdown of the pion momentum carried by quarks and gluons in $\overline{\mathrm{MS}}$ at
        $2\,\mathrm{GeV}$. Panel (a) shows the QCD JAM global PDF result of
        Ref.~\cite{Barry:2025wjx}, while panel (b) shows our large-$N_c$ result; the gluon fraction is inferred from the momentum sum rule. 
    }
    \label{fig:pion_momentum_budget}
\end{figure*}

Figure~\ref{fig:pion_momentum_budget} compares the corresponding quark--gluon
momentum balance with the $N_c=3$ JAM global PDF analysis of
Ref.~\cite{Barry:2025wjx}. At $2\,\mathrm{GeV}$, the JAM analysis gives
$\langle x\rangle_g=0.439(28)$ and therefore
$\langle x\rangle_q=0.561(28)$ from the momentum sum rule. At large $N_c$, our
AIC-averaged result $\langle x\rangle_q=0.267(105)$ corresponds to
$\langle x\rangle_g=0.733(105)$ by the same sum rule. 

\clearpage
\section{Conclusion and outlook}\label{sec:concl}

In this work, we have used the TEK model to calculate the pion quark momentum fraction directly at large $N_c$. The relevant matrix element was extracted from the response of the pion mass to a Feynman--Hellmann perturbation of the Dirac operator. Combining the results at several values of $N_c$, the lattice spacing, and the pion mass in an AIC average over correlated global fits, we obtain the renormalized momentum fraction
\begin{equation}
    \langle x\rangle_q^{\overline{\mathrm{MS}}}(2\,\mathrm{GeV})
    = 0.267(105)
\end{equation}
in the continuum limit at the physical pion mass. Our results reveal sizable finite-lattice-spacing effects, making the continuum extrapolation the dominant source of uncertainty, whereas the pion-mass dependence is milder over the simulated range. At the lattice spacing and pion mass used in the previous $N_c=17$ calculation, our AIC-averaged matched result lies approximately $1.5$ standard deviations above that determination~\cite{Karthik:2022largeNcParton}. This comparison, together with the substantial shift observed toward the continuum limit, highlights the importance of controlling finite-lattice-spacing effects.

The principal physical implication of our result is that gluons seem to carry most of the pion momentum at large $N_c$: the momentum sum rule implies
$\langle x\rangle_g^{\overline{\mathrm{MS}}}(2\,\mathrm{GeV})=0.733(105)$,
inferred from the total quark contribution. A substantial gluon contribution is consistent with standard large-$N_c$ counting. In the 't~Hooft limit with fixed $N_f$, gluon radiation from a quark remains unsuppressed, $\alpha_s C_F=O(1)$, whereas gluon splitting into additional quark--antiquark pairs is suppressed by $N_f/N_c$~\cite{tHooft:1973alw,Karthik:2022largeNcParton}. Thus, the suppression of sea quarks does not require quarks to carry most of the meson momentum. This distinction is relevant to effective descriptions such as the Nambu--Jona-Lasinio model, which admits a systematic $1/N_c$ expansion but contains no explicit gluon degrees of freedom: in its standard partonic interpretation, constituent quarks carry all of the pion momentum at a low model scale~\cite{Davidson:2001cc}.

Our continuum large-$N_c$ determination therefore provides a quantitative benchmark for descriptions of pion structure at large $N_c$ and motivates independent calculations of the gluon momentum fraction and other observables probing the structure of pions and other mesons in this limit.

\vspace{1cm}

\section*{Acknowledgements}

We thank Claudio Bonanno, Margarita Garc\'ia P\'erez and Antonio Gonz\'alez-Arroyo for useful discussions and cross-checks on the TEK results.

IZ thanks the University of Bern and the Albert Einstein Center for its hospitality, and the MISTI Switzerland program for financial support, in a summer visit during which this project was significantly advanced. 

The work of FRL was supported in part by the Swiss National Science Foundation (SNSF) through grant No. 200021-236432, and the Platform for Advanced Scientific Computing (PASC) project ``ALPENGLUE''. 

PES is supported in part by the U.S. Department of Energy, Office of Science, Office of Nuclear Physics, under grant Contract Number DE-SC0011090 and by Early Career Award DE-SC0021006, and by Simons Foundation grant 994314 (Simons Collaboration on Confinement and QCD Strings), and has benefited from the QGT Topical Collaboration
DE-SC0023646. 
This work is supported by the U.S.\ National Science Foundation under Cooperative Agreement PHY-2019786 (The NSF AI Institute for Artificial Intelligence and Fundamental Interactions, \url{http://iaifi.org/}).

Part of the analysis was performed on grants from the Swiss National Supercomputing Centre (CSCS) under project IDs \texttt{c36} and \texttt{lp53} on Alps.

Numerical experiments and data analysis used PyTorch~\cite{paszke2019pytorch}, 
Horovod~\cite{sergeev2018horovod}, 
NumPy~\cite{harris2020array}, and 
SciPy~\cite{2020SciPy-NMeth}. Figures were produced using Matplotlib~\cite{Hunter:2007}.

\appendix
\crefalias{section}{appendix}

\section{Validation of results}

\subsection{Small $\kappa$ expansion} \label{app:training}

As a check of our correlator pipeline, we
expand $D_W(p)=\mathbbm{1}-\kappa H(p)$ as
$D_W^{-1}(p)=\sum_{n\geq0}\kappa^nH^n(p)$, and compare the pion correlator 
\begin{equation}
        C_\pi(p_4) = 
        \Tr\left[
            \gamma_5
            \left(D_W(0)\right)^{-1}
            \gamma_5
            \left(D_W(p_4)\right)^{-1}
        \right]
\end{equation}
to its analytical expansion at small $\kappa$. 
After Fourier transforming to position space, the smallest time separation whose leading contribution is $o(1)$ is $t=1$, where the correlator is given (for $\sqrt{N_c} \ge 6$\footnote{In our TEK setting, if the effective lattice size $L = \sqrt{N_c}$ is less than or equal to $5$, then there are additional ``topologically nontrivial'' Wilson loops of length $\ell \le 5$ besides the usual $1 \times 1$ plaquette, which then contribute $O(\kappa^\ell)$ and invalidate the expansion as written.}) by
\begin{equation}\label{eq:small_kappa_expansion}
    C_\pi(t=1)
    =
    8N_c^2\kappa^2
    +192N_c^2\left(1+\overline W_{1,1}\right)\kappa^4
    +O(\kappa^6),
\end{equation}
where $\overline W_{1,1}$ is the average of the temporal plaquettes\footnote{The ensemble-expectation of this quantity is clearly equal to the usual plaquette expectation, but the written form is such that Eq.~\eqref{eq:small_kappa_expansion} is valid at a single-configuration level of granularity.}:
\begin{equation}
    \overline W_{1,1}
    =\frac{1}{3N_c}\sum_{i=1}^3
    \operatorname{Re}\Tr\!\left[
        z_{i4}U_4U_iU_4^\dagger U_i^\dagger
    \right].
\end{equation}

\begin{figure}[h!]
    \centering
    \includegraphics[width=\columnwidth]{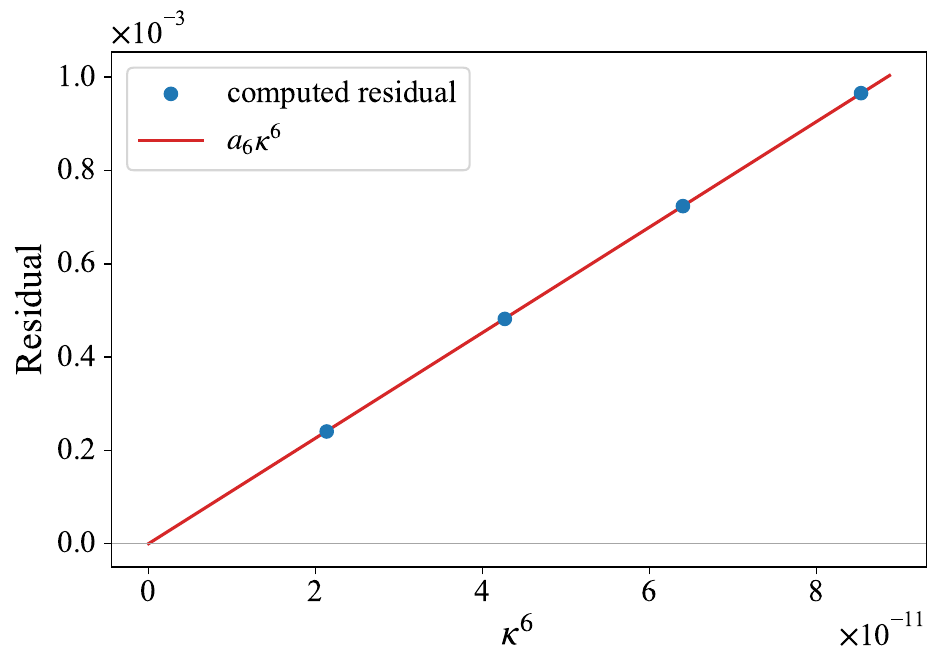}
    \caption{
        Small-$\kappa$ expansion check of the pion correlator on a
        representative interacting $N_c=36$, $k=1$ configuration. The values
        used are $\kappa=\{0.016654,0.018693,0.020000,0.020982\}$. The points show the residual after
        subtracting the $O(\kappa^2)$ and $O(\kappa^4)$ terms in
        Eq.~\eqref{eq:small_kappa_expansion}, while the line shows these residuals are well-fit by the expected $O(\kappa^6)$ behavior (with corrections at $O(\kappa^8)$).
    }
    \label{fig:hopping_expansion_checks}
\end{figure}

We test this relation on a representative interacting configuration with
$N_c=36$, $k=1$, and $b=0.36$, for which
$\overline W_{1,1}=0.5460704$. After subtracting the terms of order
$\kappa^2$ and $\kappa^4$ in Eq.~\eqref{eq:small_kappa_expansion}, the
remaining contribution follows the expected $O(\kappa^6)$ behavior, as shown
in Fig.~\ref{fig:hopping_expansion_checks}.

\subsection{Free theory checks} \label{app:free_theory}

As a validation of our Feynman--Hellmann expression Eq.~\eqref{eq:FH_mpi_deriv}, we test our numerical results in the free-field limit $b\to\infty$.
In this limit, the gauge fields are frozen to their minimum-action configuration, which in the TEK model is precisely the twist-eater configuration Eq.~\eqref{eq:twist_eater}. As the gauge fields are nondynamical, the quark momentum fraction should approach $\langle x \rangle \to 1$ in the limit as $am_\pi \to 0$.
For this check, we use untwisted Wilson fermions with $N_c=289$. In addition, we simulate a larger temporal extent  $T = 6\sqrt{N_c} = 102$ by gridding over a correspondingly finer set of temporal momenta (as discussed in Ref.~\cite{GarciaPerez:2020gnf}).

We compute the pion correlator at $\delta=0,\pm0.01$ for $\kappa=0.120,0.121,0.122,0.123,0.124$ approaching the free Wilson fermion critical value of $\kappa_c=1/8$. For each of these pairs of $(\kappa, \delta)$ we compute the pion correlator and extract the pion mass $am_\pi$ from a cosh-plus-constant fit. From these masses we compute the Feynman--Hellmann derivative
\begin{equation}
    \langle x\rangle_q^{\rm lat}
    =
    -\frac{1}{am_\pi}
    \left.\frac{\partial(am_\pi)}{\partial\delta}\right|_{\delta=0}
\end{equation}
via a symmetric finite difference. A quadratic extrapolation in $am_\pi$ gives
$\langle x\rangle_q^{\rm lat}=1$ within numerical error at $am_\pi=0$. This test directly validates the Feynman--Hellmann relation Eq.~\eqref{eq:FH_mpi_deriv} in the free-field limit.

\begin{figure*}[hb!]
    \centering
\includegraphics[width=\textwidth]{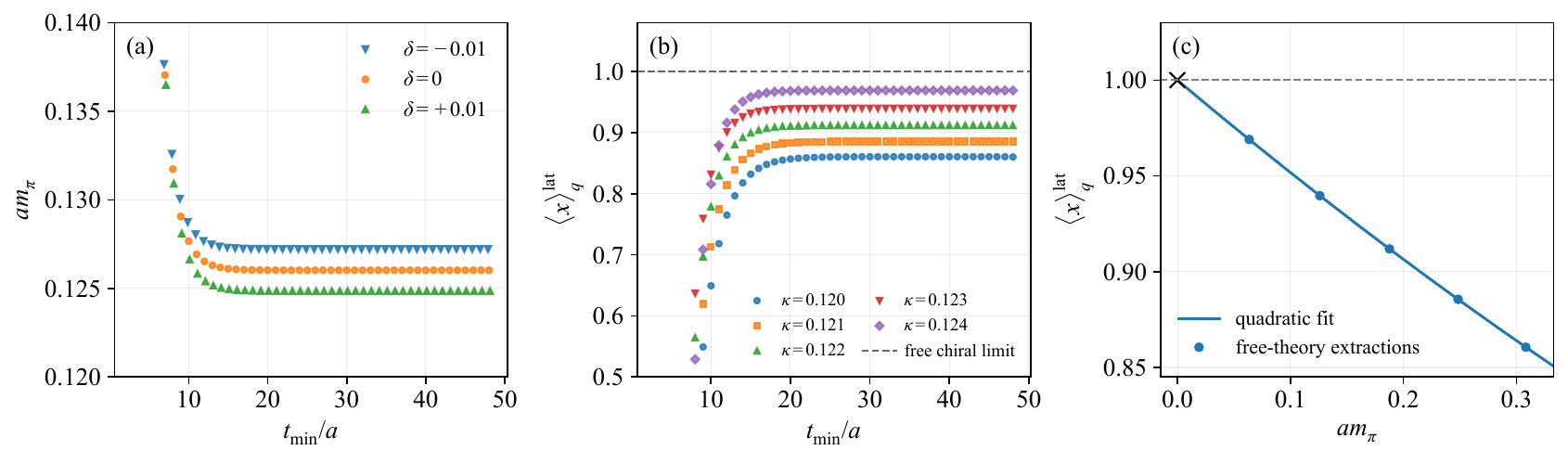}
    \caption{
        Free-field validation using untwisted Wilson fermions. Panel (a) shows
        the fitted pion mass as the left endpoint $t_{\min}$ of the fit window is varied for $\kappa=0.123$ and $\delta=0,\pm0.01$. Panel (b) shows
        the Feynman--Hellmann momentum fraction as a function of $t_{\min}$
        for all five values of $\kappa$; the dashed line denotes the free
        chiral-limit value. Panel (c) shows the extracted momentum fractions
        as functions of $am_\pi$, together with the quadratic extrapolation
        and the expected limiting point $(am_\pi,\langle x\rangle_q^{\rm lat})
        =(0,1)$.
    }
    \label{fig:free_theory_validation}
\end{figure*}

\FloatBarrier

\bibliographystyle{utphys}
\bibliography{main}

\end{document}